# Validation of a new *k-ε* model with the pressure diffusion effects in separated flows

Svetlana V. Poroseva

School of Computational Science, Florida State University, FL 32306

Gianluca Iaccarino

Center for Turbulence Research, Stanford University, CA 94305

The contribution of the "rapid" part of the pressure diffusion to the turbulent kinetic energy balance is analyzed, and a new model to describe its effect is suggested. A new transport equation for the turbulent kinetic energy is derived. The new *k*-equation does not require any modification in the standard *ε*-equation. A new *k-ε* model, which includes the new *k*-equation and the standard *ε*-equation, is validated in four separated flows: a planar diffuser, over a backstep, in a channel with wavy walls, and in an axisymmetric combustion chamber. It is shown that a new model reproduces the mean velocity, shear stress, and turbulent kinetic energy profiles and the skin friction coefficient in very good agreement with experimental data

## I. INTRODUCTION

Two-equation turbulence models (Jones and Launder[1]; Launder and Sharma[2]) are widely used in industrial CFD applications although their shortcomings are well known. Among these shortcomings are i) the limitations of the linear eddy viscosity assumption, or Boussinesq approximation, as a representation of the Reynolds stresses and ii) the difficulties in forming a transport equation for the length-scale determining quantity needed to build the eddy viscosity. Both questions have been thoroughly investigated in, e.g., Cousteix and Aupoix[3] and Apsley and Leschziner[4]. Rather than questioning either of these points, the current paper focuses on the adequacy of the standard model transport equation for the turbulent kinetic energy $k$

$$\frac{Dk}{Dt} = P - \varepsilon + \frac{\partial}{\partial x_i}\left[\left(\nu + \frac{\nu_t}{\sigma_k}\right)\frac{\partial k}{\partial x_i}\right] \tag{1}$$

in inhomogeneous turbulent flows. In equation (1), the notation is: $P = -<u_i u_j> \partial U_i/\partial x_j$, $U_i$ and $u_i$ are the mean and fluctuating velocity components, $<...>$ means ensemble average, $\nu$ is the kinematic viscosity, $\varepsilon = \nu <\partial u_i/\partial x_k \cdot \partial u_i/\partial x_k>$, $\nu_t$ is a model function representing the turbulent viscosity, and $\sigma_k$ is a model coefficient. Cartesian tensor notation is used in (1) and in what follows.

In homogeneous turbulence, the velocity – pressure-gradient correlations $<u_i \partial p/\partial x_j>$ ($p$ is the pressure fluctuation) that appear in the Reynolds averaged Navier-Stokes equation do not make any contribution to the exact transport equation for the turbulent kinetic energy. In inhomogeneous turbulent flows, however, there is a contribution of the pressure-diffusion term $\partial <u_i p>/\partial x_i$ to the turbulent kinetic energy balance. Direct numerical simulation (DNS) data from free shear flows (Rogers and Moser[5], Moser et al.[6]) show that the contribution of the pressure diffusion to the turbulent kinetic energy balance is not negligible, especially in the central core of these flows. These data also show that modeling the pressure diffusion and turbulent diffusion terms together is not likely to be successful because they have qualitatively different profiles (see Fig. 21 in Rogers and Moser[5] and Fig. 12 in Moser et al.[6]). In fact, the model $\partial(v_t/\sigma_k \cdot \partial k/\partial x_i)/\partial x_i$ for the turbulent diffusion in equation (1) can absorb only a part of the pressure-diffusion term $<u_i \partial p/\partial x_i>$, the so-called "slow" part:

$$-\frac{1}{\rho}<u_i \frac{\partial p}{\partial x_i}>^{(s)} = \frac{1}{5}\frac{\partial <u_m u_m u_i>}{\partial x_i} \qquad (2)$$

(Lumley[7]). In (2), $\rho$ is the density and $\partial <u_m u_m u_i>/\partial x_i$ is the turbulent diffusion. Thus, the "rapid" pressure-diffusion term does not appear in equation (1) and requires modeling.

Evidence that important information is missing in equation (1) is that two-equation models employing this equation can generally reproduce the mean velocity and shear stress profiles accurately, but the turbulent kinetic energy level is not well predicted, even in the central core of free shear flows (Poroseva[8]; Poroseva and Bézard[9]) where wall effects are irrelevant. Notice also that the transport equation for the dissipation $\varepsilon$ (or any other second scale) is formally derived from the k-equation. Therefore, any physics missing in the k-equation causes the loss of accuracy in describing the second scale transport regardless the choice for this scale.

It was shown by Poroseva[8] that the turbulent kinetic energy level could be corrected if the "rapid" part of the pressure diffusion term (related to the mean velocity gradients) is included in the k-equation as an extra term related to the production P. The contribution of the pressure diffusion to the transport equation for $\varepsilon$ ($\varepsilon$-equation) appears through one of the modeling coefficients in the $\varepsilon$-equation. A two-equation model with the new k-equation was validated in Poroseva[8] in self-similar free shear flows (plane wake, plane mixing layer, plane and round jets) and equilibrium boundary layers with and without pressure gradients. The objective of the current paper is to investigate further potential of the model in application to simulations of the separated flows in a planar diffuser, over a backstep, in a channel with wavy walls, and in an axisymmetric combustion chamber (Fig. 1).

**II. MODELING THE PRESSURE DIFFUSION**

Excluding from the consideration a flow area very close to the flow boundary, the pressure diffusion correlation in the exact transport equation for the turbulent kinetic energy can be presented as a sum of "rapid" and "slow" parts

$$-\frac{1}{\rho}\frac{\partial <u_i p>}{\partial x_i} = -\frac{1}{\rho}(\frac{\partial <u_i p>^{(r)}}{\partial x_i} + \frac{\partial <u_i p>^{(s)}}{\partial x_i}) \quad (3)$$

(Chou[10]). The second term on the right hand side of (3) can be modeled by expression (2) as discussed above. Therefore, its contribution in the turbulent kinetic energy balance is absorbed by an adopted model for the turbulent diffusion. The first term on the right hand side of (3) is called the "rapid" part due to its relation to the mean velocity gradient. A model for this term

$$-\frac{1}{\rho}\frac{\partial <u_i p>^{(r)}}{\partial x_i} = \left(-\frac{3}{5} + C_k\right)P, \quad (4)$$

was first suggested in Poroseva[8]. Expression (4) is obtained by analyzing the properties of the exact integral expression for the "rapid" part of the velocity – pressure-gradient correlation in an incompressible flow

$$-\frac{1}{\rho}<u_i \frac{\partial p}{\partial x_j}>^{(r)} = -\frac{1}{2\pi}\iiint \frac{\partial}{\partial x'_j}\left[\frac{\partial U'_m}{\partial x'_n}\frac{\partial <u'_n u_i>}{\partial x'_m}\right]\frac{1}{r}dV' \quad . \quad (5)$$

In (5), " ' " above a flow variable indicates that it should be evaluated at a point $Y'$ with coordinates $x'_i$, which ranges over the region of the flow; $r$ is the distance from $Y'$ to the point $Y$ with coordinates $x_i$; $dV'$ is the volume element. The velocity – pressure-gradient correlation on the left side of (5) is evaluated at point $Y$, whereas all derivatives on the right side are taken at $Y'$. The analysis yields the following model for this correlation

$$-\frac{1}{\rho}<u_i p_{,j}>^{(r)} = a_{nmji}U_{m,n}, \quad (6)$$

where

$$a_{nmji} = -\frac{1}{5}\left(<u_i u_j>\delta_{mn} + <u_i u_m>\delta_{jn}\right) + \frac{4}{5}<u_i u_n>\delta_{jm} +$$
$$C_1\left[\frac{1}{2}\left(<u_i u_j>\delta_{mn} + <u_i u_m>\delta_{jn}\right) + k\left(\delta_{ij}\delta_{mn} + \delta_{im}\delta_{jn}\right) + <u_i u_n>\delta_{jm}\right.$$
$$\left. - <u_j u_m>\delta_{in} - 2\left(<u_j u_n>\delta_{im} + <u_m u_n>\delta_{ij}\right)\right] + \quad (7)$$
$$C_2\left[\frac{1}{2}\left(<u_i u_j>\delta_{mn} + <u_i u_m>\delta_{jn} - <u_j u_n>\delta_{im} - <u_m u_n>\delta_{ij}\right)\right.$$
$$\left. + k\delta_{in}\delta_{jm} - \frac{3}{2}<u_j u_m>\delta_{in}\right].$$

In expression (7), $\delta_{ij}$ is the Kronecker symbol; the model coefficients $C_1$ and $C_2$ are generally unknown functions of several parameters. Derivation of expressions (6) and (7) is described in detail in Poroseva[11]. Notice that these expressions are derived without the assumption of homogeneity.

In the transport equation for the turbulent kinetic energy, expression (6) with model (7) for the tensor function $a_{nmji}$ contracts to (4), where

$$C_k = \frac{15}{2}\cdot C_1 + 3C_2. \quad (8)$$

In general, the coefficient $C_k$ is a function of the same parameters as the coefficients $C_1$ and $C_2$. In homogeneous turbulence, however, the coefficients $C_1$ and $C_2$ are linked:

$$\frac{1}{5} - \frac{5}{2}C_1 - C_2 = 0 \quad (9)$$

(Poroseva[11]). Substitution of (9) in expression (8) yields the universal constant value of $C_k$ equal to $0.6$. That is, in homogeneous turbulence, the pressure diffusion term does not contribute to the turbulent kinetic energy balance as expected.

It should be pointed out that model expression (4) for the "rapid" part of the pressure diffusion is similar to the model derived *ad hoc* by Demuren *et al.*[12] by analyzing direct numerical simulation (DNS) data.

An important question in modeling $<u_i p>_{,i}$ is whether a model expression for this term should be of the diffusive type. In regard to the "rapid" part of the correlation $<u_i p_{,j}>$ (see expression (5)), there is no indication that the model for this term should be of the diffusive type. What "diffusive type" requires is that the integral of the sum of three correlations $<u_i p>_{,i}$ ($i=1,2,3$) taken over the entire flow volume vanishes. It does not imply that the sum of three correlations $<u_i p>_{,i}$ vanishes at every point in the flow, or that any one of $<u_1 p>_{,1}$, $<u_2 p>_{,2}$, and $<u_3 p>_{,3}$ vanishes throughout the flow. This requirement does not also imply that each of the terms in expression (3) (which in its complete form includes boundary effects) would vanish separately. Finally, even assuming that the integral of (4) taken over the entire flow volume should vanish, one can argue that this result can be achieved with different functional forms of the coefficient $C_k$, not necessarily of the diffusive type. (More discussion on a functional form of the model coefficients is provided in Section IV.) This question clearly requires more study in the future. Notice, however, that expressions (4) and (7) hold regardless of the models for the coefficients $C_1$, $C_2$, and $C_k$.

## III. TURBULENCE MODEL

Using model expression (4) for the "rapid" part of the pressure diffusion, one obtains a new model equation for the turbulent kinetic energy transport:

$$\frac{Dk}{Dt} = (0.4 + C_k)P - \varepsilon + \frac{\partial}{\partial x_i}\left[\left(\nu + \frac{\nu_t}{\sigma_k}\right)\frac{\partial k}{\partial x_i}\right]. \tag{10}$$

In homogeneous turbulence with $C_k = 0.6$, equation (10) transforms to the standard *k*-equation (equation (1)).

The form of equation (10) does not require any modification in the standard $\varepsilon$-equation

$$\frac{D\varepsilon}{Dt} = \frac{\varepsilon}{k}(C_{\varepsilon 1} P - C_{\varepsilon 2} \varepsilon) + \frac{\partial}{\partial x_i}\left[\left(\nu + \frac{\nu_t}{\sigma_\varepsilon}\right)\frac{\partial \varepsilon}{\partial x_i}\right], \tag{11}$$

where $C_{\varepsilon 1}$, $C_{\varepsilon 2}$, and $\sigma_\varepsilon$ are model coefficients. The effect of the "rapid" part of the pressure diffusion comes from the coefficient $C_{\varepsilon 1}$, which generally should be a function of the same parameters as the coefficient $C_k$. However, since a functional form is currently unavailable for both coefficients, we approximate them in this study by constant values and then, investigate whether this approximation results in improvement of simulation results.

Indeed, if one allows the coefficient $C_{\varepsilon 1}$ being constant in any given flow, but variable from flow to flow, even the standard $k-\varepsilon$ model (equations (1) and (11)) reproduces the mean velocity and shear stress profiles in a good agreement with experimental data in free shear flows and equilibrium boundary layers under different pressure gradients (Poroseva and Bézard[9]). The results of simulations obtained with $C_{\varepsilon 1}$ variable from flow to flow (see Table 2 in Poroseva and Bézard[9]) and the rest of the coefficients given by

$$\nu_t = C_\mu k^2/\varepsilon, \, C_\mu = 0.09, \, C_{\varepsilon 2} = 1.92, \, \sigma_\varepsilon/\sigma_k = 1/0.67 = 1.5 \tag{12}$$

are better than the results obtained with the $k$-$\omega$ (Wilcox[13]) and $k$-$\varphi$ (Cousteix et al.[14]) models. However, the axis level of the turbulent kinetic energy is either overestimated (plane wake) or underestimated (plane jet, mixing layer, boundary layers). Only in the round jet, the turbulent kinetic energy level is reproduced correctly with $C_{\varepsilon 1} = 1.5$ (see Fig.5 in Poroseva and Bézard[9]). It is interesting to note that the same value $C_{\varepsilon 1} = 1.5$ was recommended as optimal for homogeneous flows also (Kassinos et al.[15]).

Allowing the coefficient $C_k$ deviate from its homogeneous value 0.6 and both coefficients $C_k$ and $C_{\varepsilon 1}$ vary from flow to flow as shown in Table 1, one can reproduce well not only the mean velocity and shear stress profiles, but also the turbulent kinetic energy level in the plane wake, the plane jet, and the mixing layer, as well as in the equilibrium boundary layers (Poroseva[8]).

In Section IV, the $k-\varepsilon$ model including equations (10) and (11) with the coefficients $C_k$ and $C_{\varepsilon 1}$ being constant, but variable from flow to flow, and with the rest of the model coefficients given by (12) is validated in four separated flows: a planar diffuser, over a backstep, in a channel with wavy walls, and in an axisymmetric combustion chamber (Fig. 1). We will denote this model as LS-RPD. The performance of the LS-RPD model will be compared with the performance of the standard $k-\varepsilon$ model (Launder and Sharma[2]). This model (hereinafter referred as LS) includes equations (1) and (11), with the model coefficients being set to their standard values: $C_{\varepsilon 1}=1.44$, $C_{\varepsilon 2}=1.92$, $C_\mu=0.09$, and $\sigma_\varepsilon/\sigma_k=1.3/1=1.3$. Results obtained using the four-equation $<v^2>-f$ model (Durbin[16]) are also included for comparison.

Equations (1), (10), and (11) are written in the high Reynolds number form. Since, modeling wall effects (or low Reynolds number effects) is not in the focus of the current paper, the standard damping function approach proposed by Launder and Sharma[2] is used to correct the behavior of turbulent quantities in the viscous dominated near-wall regions. Steady equations are solved using the commercial code Fluent.

## IV. RESULTS AND DISCUSSION

The first problem selected is the backstep flow (Jovic and Driver[17]) shown in Fig.1(a). The Reynolds number based on the inlet velocity ($u_i$) and the step height (*H*) is 5,100. The flow at the inlet is a fully developed boundary layer. Separation is fixed at the step and the expansion generates a large recirculating region with strong negative velocity and high turbulent kinetic energy (measurements are available at several stations downstream the step). The coefficients $C_k$ and $C_{\varepsilon 1}$ for the model LS-RPD are given in Table 2. Their values are chosen such as to fit the experimental data. Figure 2 shows the experimental and calculated profiles of the streamwise velocity and the turbulent kinetic energy (both scaled by the inlet velocity) at three spatial locations in the streamwise direction *x*; *y* is the vertical direction. All three models – LS, LS-RPD, and $<v^2>-f$ – give the same mean velocity profile upstream ($x/H=-3$) of the backstep (Fig. 2(a)). The level of the turbulent kinetic energy at this location is overpredicted by the LS-RPD model and underpredicted by the LS model. On this and other figures, dashed lines show the profiles calculated with the LS model, solid lines show the LS-RPD model profiles, and the dash-double-dotted lines correspond to the results produced with the $<v^2>-f$ model. Experimental data are given by white circles.

Downstream, the LS model does not correctly reproduce the separation zone at both locations ($x/H = 4$ and 6), whereas other two models are in a good agreement with the experimental data. In addition, the LS model fails to calculate the correct friction coefficient $c_f$ in the recirculating bubble and underestimates $c_f$ in the recovery region, whilst the LS-RPD and $<v^2>-f$ models produce similar friction levels (Fig. 3)

The second test case is the flow in the asymmetric diffuser shown in Fig. 1(b). The flow is fully developed at the inlet. The Reynolds number based on the bulk velocity ($u_i$) and the inlet height (*H*) is 20,000. The presence of a mild adverse pressure gradient induces a separation on a smooth surface, which is very challenging for turbulence models. Mean velocity and turbulent kinetic energy profiles are available as well as wall skin friction (Buice and Eaton[18]) to identify the extent of the separated region. As in the previous case, the LS results are in a poor agreement with the measurements (Fig. 4) at three streamwise locations ($x/H = 24, 28, 32$). Both LS-RPD (with the coefficients $C_k$ and $C_{\varepsilon 1}$ shown in Table 2) and $<v^2>-f$ models produce results, which are comparable in accuracy for the mean velocity and the turbulent kinetic energy at all three locations. These results are in a good agreement with the experimental data. Also, both models capture the extent of the separation region very well (Fig. 5).

The third case is the flow in a periodic wavy channel (Fig. 1(c)). The Reynolds number based on the bulk velocity ($u_i$) and the average channel height (*H*) is 11,000. The flow separates on the downhill slope and reattaches on the uphill. Only velocity measurements are available in this case (Kuzan[19]). Since the level of the turbulent kinetic energy is not known in this case, the value of $C_k$ cannot be chosen and is set to 0.6, its homogeneous value. The value of $C_{\varepsilon 1}$ is chosen to fit the experimental data for the mean velocity (see Table 2)). As Figure 6 demonstrates, three models produce reasonably accurate results at two spatial locations $x/H = 0.25$ and 0.75.

The fourth case consists of the axisymmetric combustion chamber (Fig.1(d)). A central pipe stream and an annular swirling stream enter a large cylindrical chamber, and in response to a strong adverse pressure gradient, a recirculating region is created. The Reynolds number based on the pipe bulk velocity ($u_i$) and diameter ($=2R$) is 75,000. Streamwise ($u$) and swirl velocities ($w$) are measured at three stations ($x/R = 0.7, 1.68, 3.6$) in the chamber (Hagiwara et al.[20]) Again, because experimental data is not available for the turbulent kinetic energy, the coefficient $C_k$ in the LS-RDT model is set (not chosen) to 0.6. The value of $C_{\varepsilon 1}$ is given in Table 2. The LS model considerably overestimates the extent of the recirculating bubble and reproduces poorly the streamwise and swirl velocities at the location $x/R = 1.68$ (Fig. 7 (b)). At other locations, the performance of three models is comparable in accuracy and is in good agreement with the experimental data.

The results reported in this paper and Poroseva[8] demonstrate that the approximation of the coefficients $C_k$ and $C_{\varepsilon 1}$ by constants works well in all test flows with different geometries and at different Reynolds numbers provided that the values of these coefficients are allowed to change from flow to flow. However, to use equations (10) and (11) for predictions, that is, in simulation of flows for which experimental data are not available, a general functional form should be found for $C_k$ and $C_{\varepsilon 1}$. Currently, only few conclusions about the coefficients $C_k$ and $C_{\varepsilon 1}$ can be drawn based on available experimental and DNS data.

In each test flow, the mean velocity and shear stress profiles (if available) appear to be insensitive to the value of the coefficient $C_k$. That is, they can be well reproduced with $C_k = 0.6$ corresponding to homogeneous turbulence and $C_{\varepsilon 1}$ chosen to fit the experimental mean velocity profile. (The rest of coefficients is set to (12).) Interesting enough, in the round jet and diffuser flows, the optimal value found for $C_{\varepsilon 1}$ ($C_{\varepsilon 1} = 1.5$) is the same as recommended for homogeneous turbulence in Kassinos et al.[15]. What determines the final choice of the set ($C_k$, $C_{\varepsilon 1}$) in a given flow is the turbulent kinetic energy level controlled by the coefficient $C_k$. It appears, however, that in geometrically-equivalent flow situations at the same Reynolds number, multiple asymptotic states can be observed (see discussion in Rogers and Moser[5] and Moser et al.[6]) That is, whereas the mean velocity and shear stress profiles are universal (or nearly universal) under appropriate scaling, the normal stresses and turbulent kinetic energy profiles are non-unique. DNS confirms that multiple asymptotic states reflect the differences in the large-scale structure of turbulence, which depends strongly on the Reynolds number, "uncontrolled and possibly unknown properties of the initial or inlet conditions" (Moser et al.[6]), flow geometry, boundary conditions, external forces etc. (Tsinober[21]). Since this is the coefficient $C_k$, which controls the level of the turbulent kinetic energy, the large-scale structure should be reflected through this coefficient.

The coefficient $C_k$ is linked to the coefficients $C_1$ and $C_2$ through expression (8). Thus, $C_k$ should be a function of the same parameters as these two coefficients. Considering the limiting states of turbulence (Poroseva[11]), one can show that $C_1$ and $C_2$ are functions at least of the mean velocity gradients and the Reynolds stresses even in homogeneous turbulence. No universal constant values exist for $C_1$ and $C_2$. Nevertheless, the coefficient $C_k$ does have such a value in homogeneous turbulence: $C_k = 0.6$. This fact clearly indicates that $C_1$ and $C_2$ should also depend on parameters which directly relate to inhomogeneous effects such as, e.g., second derivatives of flow characteristics and other parameters not specified yet. More study (including DNS) is necessary to determine general functional forms for the coefficients $C_1$, $C_2$, and $C_k$.

As for the coefficient $C_{\varepsilon 1}$, its value in different flows is close to each other (see Tables 1 and 2). However, no single value can be recommended for all considered flows. Since variability of both coefficients $C_k$ and $C_{\varepsilon 1}$ is linked to the same mechanism, that is, to the pressure diffusion effects, functional forms for both coefficients cannot be considered separately.

## V. SUMMARY

In the current paper, the contribution of the "rapid" part of the pressure diffusion to the turbulent kinetic energy balance was modeled. In the transport equation, the effect of the "rapid" part of the pressure diffusion manifests itself through an additional term related to the production term. This term contains the model coefficient $C_k$, which is generally a function of unknown parameters related to the large-scale structure of turbulence and inhomogeneous effects. In homogeneous turbulence, the coefficient takes the universal value: $C_k = 0.6$.

A new $k - \varepsilon$ model with the additional term linked to the pressure diffusion effects in the transport equation for the turbulent kinetic energy was derived. The standard $\varepsilon$-equation does not require any modification. The pressure diffusion effects influence this equation through the coefficient $C_{\varepsilon 1}$, which is a counterpart of the coefficient $C_k$ in the $k$-equation. The new model was validated in four separated flows: a planar diffuser, over a backstep, in a channel with wavy walls, and in an axisymmetric combustion chamber. The results obtained in these flows complement the results for free shear flows and equilibrium boundary layers reported previously in Poroseva[8].

Since a general functional form is currently unavailable for both coefficients $C_k$ and $C_{\varepsilon 1}$, they were approximated by constant values in simulations of all test flows. The general conclusion is that a very good agreement between experimental and calculated profiles of the mean velocity, the shear stress, and the turbulent kinetic energy as well as the skin friction coefficient can be achieved if one sets these two coefficients to be constant in any given flow, but variable from flow to flow. Variability in values of the coefficients can be linked, at least partially, to the large-scale turbulence structure. Available data, however, are not sufficient to understand and describe this connection between the model coefficients and the large-scale structure in a general form.

The results of simulations of the separated flows with the new $k-\varepsilon$ model are much better than the results produced with the standard $k-\varepsilon$ model and comparable in accuracy with the computational results of more complex four-equation $<v^2>-f$ model.

The proposed form of the $k$-equation is not complete in a sense that model expressions used in the equation to represent other terms are not consistent with each other and with the models for both parts of the pressure diffusion term. "Consistency" of models for different terms means that all models are derived based on the same assumptions and approximations. In this sense, however, no two-equation model, which can be called "complete", is currently available. This issue should be addressed in future studies.

**ACKNOWLEDGMENTS**

Dr. Svetlana V. Poroseva conducted a part of this research when was affiliated with the Center for Turbulence Research (Stanford University). The first author would also like to thank Robert Rubinstein (NASA-Langley Research Center) and M. Y. Hussaini (School of Computational Science, Florida State University) for support in the preparation of this work for publication.

TABLE 1. The value of the model coefficients $C_k$ and $C_{\varepsilon 1}$ in free shear flows and equilibrium boundary layers ($\beta$ is the pressure gradient parameter).

| flow | wake | mixing layer | plane jet | round jet | boundary layers $\beta = 19.6$ | $\beta = 0$ |
|---|---|---|---|---|---|---|
| $C_{\varepsilon 1}$ | 0.6 | 1.9 | 2.12 | 1.5 | 1.85 | 2.2 |
| $C_k$ | 0.2 | 0.9 | 1 | 0.6 | 0.8 | 1 |

TABLE 2. Values of the coefficients $C_k$ and $C_{\varepsilon 1}$ in separated flows shown in Fig.1. The values in the parenthesis are set, not chosen.

| flow | backstep | diffuser | wavy channel | combustion chamber |
|---|---|---|---|---|
| $C_{\varepsilon 1}$ | 1.85 | 1.5 | 1.5 | 1.7 |
| $C_k$ | 0.8 | 0.6 | (0.6) | (0.6) |

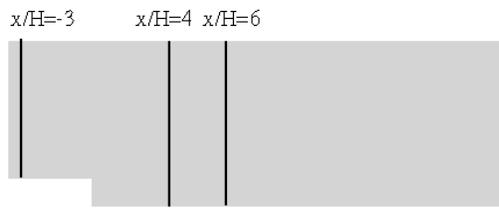
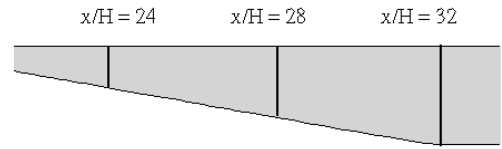
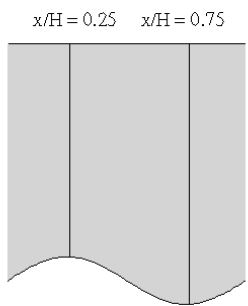
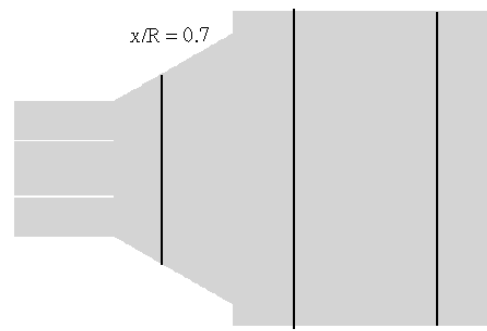

**Figure 1** Test flows: (a) backstep flow, (b) diffuser, (c) wavy channel, (d) combustion chamber (black vertical lines show the measurement locations).

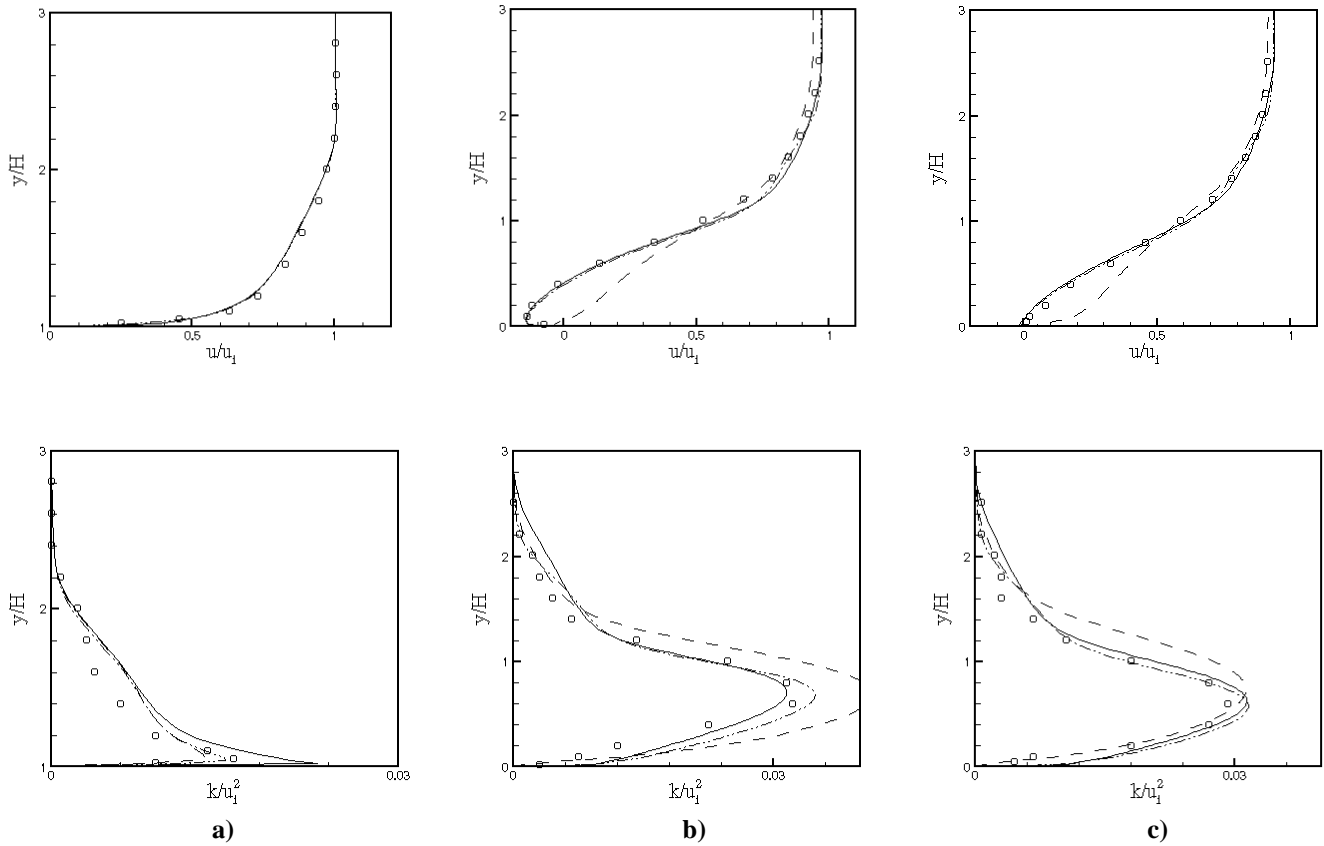

**Figure 2** Backstep flow: velocity (top) and turbulent velocity energy (bottom) profiles. Locations: (a) $x/H = -3$, (b) $x/H = 4$, (c) $x/H = 6$. Notation: experimental data (circles), the LS-RPD model (solid lines), the LS model (dashed lines), $<v^2>-f$ model (dash-double-dotted lines).

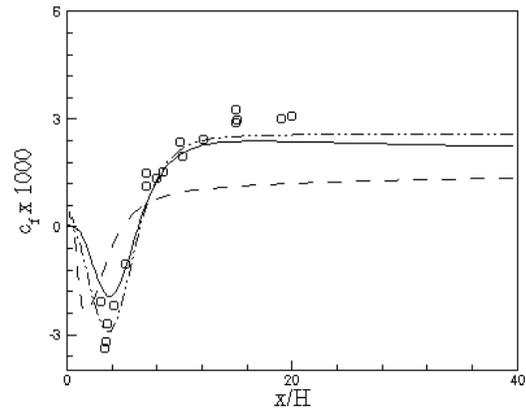

**Figure 3** Backstep flow: skin friction coefficients. (See notation on Fig. 2.)

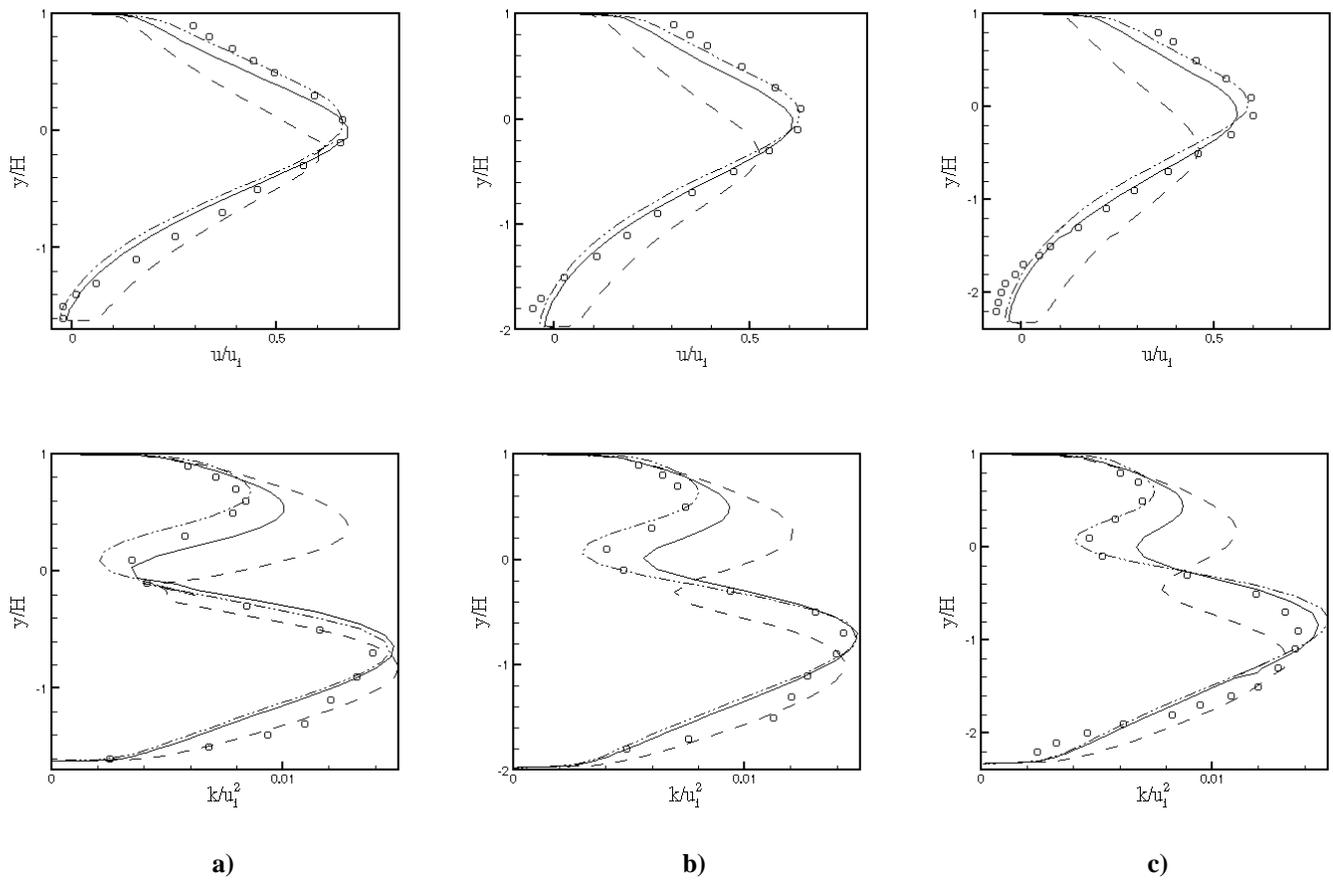

**Figure 4** Diffuser: velocity (top) and turbulent velocity energy (bottom) profiles. Locations: (a) $x/H = 24$, (b) $x/H = 28$, (c) $x/H = 32$. (See notation on Fig. 2.)

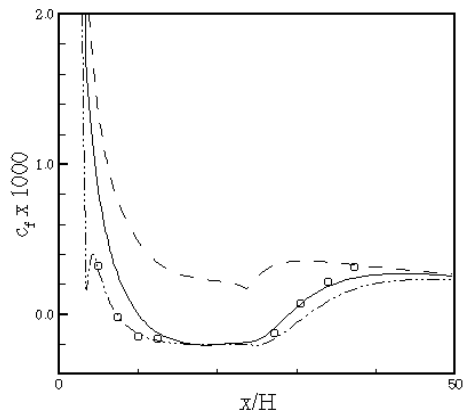

**Figure 5** Diffuser: skin friction coefficients. (See notation on Fig. 2.)

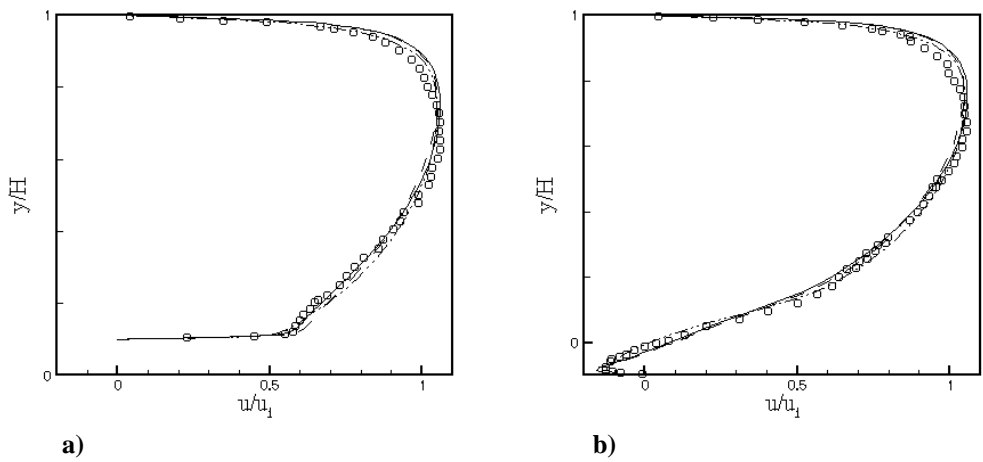

**Figure 6** Wavy channel: axial velocity profiles. Locations: (a) $x/H = 0.25$, (b) $x/H = 0.75$. (See notation on Fig. 2.)

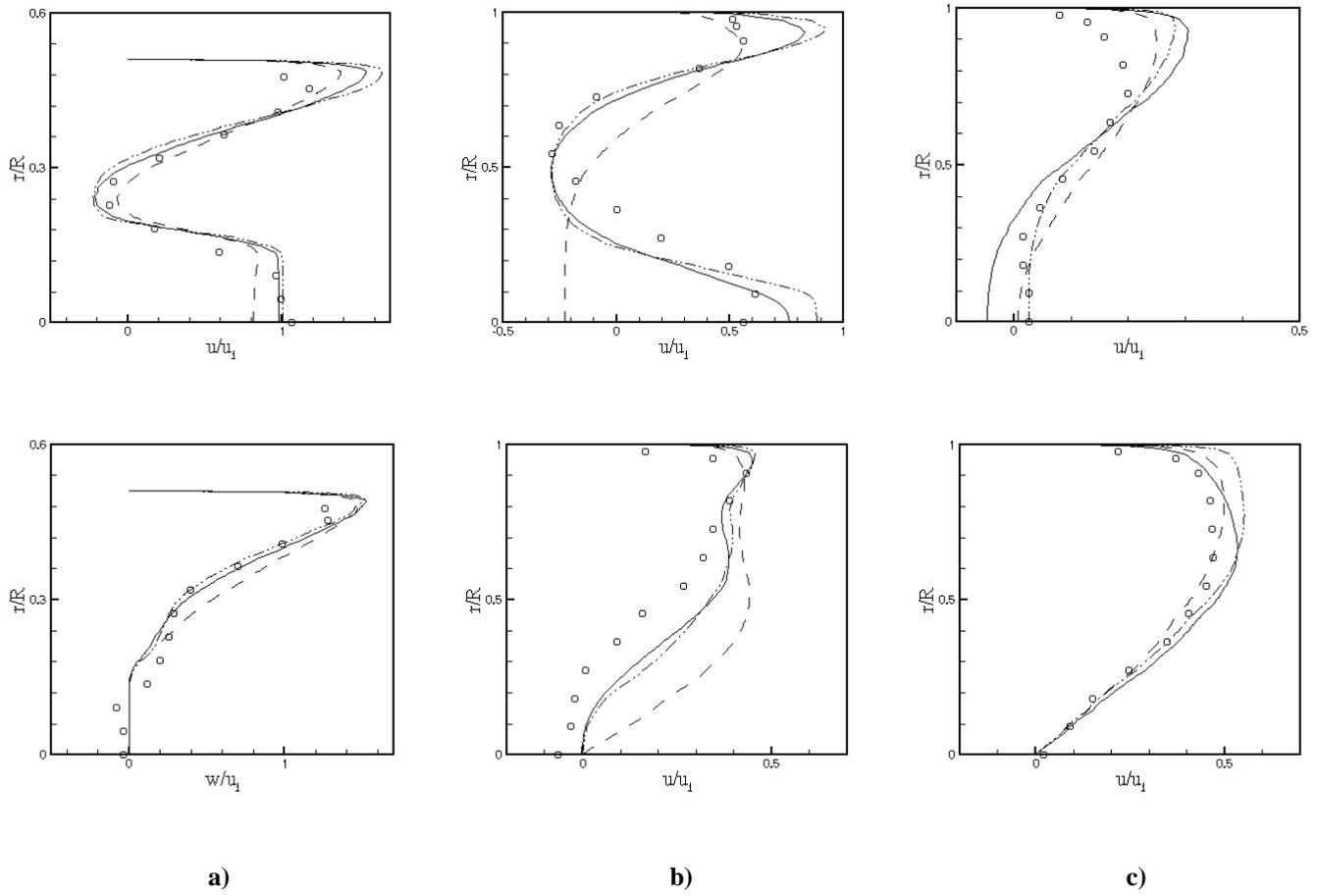

**Figure 7** Combustion chamber: axial (top) and swirl (bottom) velocity profiles. Locations: (a) $r/R = 0.7$, (b) $r/R = 1.68$, (c) $r/R = 3.6$. (See notation on Fig. 2.)